\documentclass[fleqn,usenatbib]{mnras}

\usepackage[T1]{fontenc}
\usepackage{ae,aecompl}

\usepackage{graphicx}
\usepackage{amsmath}
\usepackage{amssymb}

\title[Numerical simulations of polarisation in gamma-ray burst afterglows]{Numerical simulations of polarisation in gamma-ray burst afterglows}

\author[Medina Covarrubias et al.]{
Rogelio Medina Covarrubias$^{1}$\thanks{E-mail: rogelio.medina@correo.nucleares.unam.mx},
Fabio De Colle$^{1}$, Gerardo Urrutia$^{2}$ and Felipe Vargas$^{1}$
\\
$^{1}$Instituto de Ciencias Nucleares, Universidad Nacional Aut\'onoma de M\'exico, A. P. 70-543 04510 D. F. Mexico\\
$^{2}$Center for Theoretical Physics, Polish Academy of Sciences, Al. Lotnikow 32/46  02-668 Warsaw, Poland
}

\date{Accepted XXX. Received YYY; in original form ZZZ}

\pubyear{2023}

\begin{document}
\label{firstpage}
\pagerange{\pageref{firstpage}--\pageref{lastpage}}
\maketitle

\begin{abstract}
We compute the linear polarisation during the afterglow phase of gamma-ray bursts, for both on-axis and off-axis observers. We use numerical simulations of the deceleration of a relativistic jet, and compute the polarisation by post-processing the results of the numerical simulations. In our simulations, we consider a magnetic field that is chaotic in the plane of the shock, in addition to  a magnetic field component that is parallel to the shock velocity.  While the linear polarisation computed for on-axis observers is consistent with previous analytical estimates, we found that lateral expansion, which is accurately handled in our simulations, plays a crucial role in determining the linear polarisation for off-axis observers. Our results show that the off-axis linear polarisation, as seen by off-axis observers, exhibits a single peak, in contrast to the two peaks inferred by previous analytical studies. The maximum polarisation degree is 40\% at an observing angle $\theta_{\rm obs}=0.4$ rad, and it decreases as the observing angle increases, which is opposite to what predicted by analytical models, where polarisation increases with larger observing angles.
From the upper limit of 12\% in the linear polarisation obtained at 244 days for the GRB 170817A, we also infer an anisotropy factor of $B_\parallel/B_\perp = 0.5-0.9$, consistent with the post-shock magnetic field being amplified by turbulence.
\end{abstract}



\section{Introduction}

Gamma Ray Bursts (GRBs) are intense pulses of gamma rays emitted during the propagation of collimated relativistic jets. These jets are formed during the collapse of massive stars or the merging of compact objects, such as a neutron star or a neutron star-black hole  binary system.
The high energy prompt emission is typically followed by a multi-wavelength afterglow emission that spans from radio to X-rays frequencies  (see, e.g., \citealt{kumar15, levan18} for reviews).

The afterglow emission is accurately modelled by considering synchrotron emission produced by non-thermal electrons accelerated at the shock front and moving through the magnetised post-shock region \citep[e.g.,][]{rees92, paczynski93, meszaros97, sari98}. Thus, studying the afterglow emission is important to understand the interaction between the relativistic jet and the surrounding environment. In the best-case scenario, in which a detailed spectrum as a function of time is observed, it becomes possible to determine the density of the ambient medium, the energy of the explosion, and the micro-physical parameters of the particle acceleration process \citep[see, e.g.,][and references therein]{aksulu20, aksulu22}.

The origin and structure of the magnetic field in the post-shock region are still uncertain. The two-stream, relativistic Weibel instability can generate magnetic fields close to equipartition \citep[][]{medvedev99}, mainly tangled in the plane of the shock. However, this magnetic field will decay unless a dynamo generated by 
turbulence amplifies the magnetic field in the bulk of the post-shock region where most of the radiation is emitted \citep[e.g., ][]{milosavljevic06, sironi07, goodman08}, amplifying the component of the magnetic field perpendicular to the shock plane\citep[][]{gruzinov99a,granot03}.

As polarisation depends on the magnetic field geometry, modelling existing polarisation observations can provide information on the magnetic field orientation and origin. Different magnetic field geometries have been considered, including ordered magnetic fields, \citep[e.g.,][]{granot03a, granot03, lyutikov03, Mundell13, nakar03, cheng20, Teboul&Shaviv2021, kuwata23}, magnetic fields tangled in the shock plane or perpendicular to it \citep{ghisellini99, gruzinov99a, medvedev99, sari99, granot02}, or asymmetric magnetic fields, e.g., those generated by causally disconnected regions \citep[e.g.,][]{gruzinov99b}.

Ordered magnetic fields are known to exhibit a high level of polarisation degree (PD hereafter). \citet{laing80} demonstrated that a chaotic magnetic field, compressed along the direction of propagation of the shock front, increases the polarisation degree when the line of sight is parallel to the direction of compression. Hence, a low PD (e.g., $\lesssim 10\%$) could be an indicator of a tangled magnetic field in the plane of the shock or a magnetic field perpendicular to the shock plane. Understanding the magnetic field structure can provide insights into the particle acceleration process, jet structure, and the dynamics of the magnetic field, such as the presence of turbulence in the post-shock region.

During the early afterglow, the optical PD can reach values of $\gtrsim 10\%$ and is believed to be associated with the propagation of the reverse shock and the presence of ordered magnetic fields (see \citealt{jordana-mitjans20} and references therein). However, the optical and radio afterglow exhibit low or no polarisation ($\lesssim$ a few $\%$) at late times ($\gtrsim 1$ day) when the forward shock, which decelerates as it propagates through the ambient medium, dominates the light curve \citep[e.g.,][]{Covino&Gotz16}.

The polarisation of GRBs during the late afterglow phase has been studied extensively \citep[e.g.][]{sari99, Klose2004, wu05, Toma2008, LanWuDai15, Nava2016, gill18, LanWuDai2018, Gill&Granot2020, Birenbaum&Bromberg2021, Shimoda&Toma2021, Teboul&Shaviv2021}. Previous models typically rely on simple analytical or semi-analytical descriptions of the dynamics of the emitting region, such as a decelerating thin shell, without a radial structure (but see \citealt{Gill&Granot2020}) or lateral expansion, or with a lateral expansion described by a simple analytical prescription (i.e. by considering a jet expanding laterally at the local sound speed), and radial velocities. At late times, when the shock Lorentz factor becomes $\Gamma_{\rm sh} \lesssim 1/\theta_j$ (being $\theta_j$ the jet opening angle), the lateral expansion becomes important and should be taken into account. Furthermore, turbulence and shearing, which can only be captured via numerical calculations, can affect the velocity and magnetic field structure in the post-shock region. 

In this paper, we extend previous results by presenting the first numerical simulations of polarisation in GRB afterglows. We post-process the results of special relativistic hydrodynamic simulations that properly include jet lateral expansion. Our simulations consider magnetic fields parallel and perpendicular to the shock front, for on-axis and off-axis observers, for a top-hat jet decelerating in a uniform medium. The study of structured jets is left for future study.

This paper is structured as follows. In section 2, we describe the numerical simulations and the methods used to compute synchrotron radiation and the resulting polarisation as a function of magnetic field structure. In section 3, we present polarisation and position angles computed for different magnetic field configurations. In section 4, we discuss the results in terms of available observations. Finally, we provide our conclusions in section 5.

\section{Methods}
\label{sec:methods}

\subsection{Hydrodynamics simulations}

We compute the evolution of the PD during the afterglow phase of GRBs by post-processing the results of two-dimensional (2D), axisymmetric simulations. The simulations were performed using the adaptive mesh refinement code \emph{Mezcal}  \citep{decolle12a}, which solves the special relativistic hydrodynamic equations.

The simulations follow the deceleration of a shock with an initial Lorentz factor $\Gamma_{\rm sh} = 20\sqrt{2}$ and an opening angle $\theta_{\rm j} = 0.2$ rad. 
The post-shock density, velocity, and pressure of the relativistic jet are initialised using the self-similar solution by \citet{blandford76}, which describes the deceleration of a spherical relativistic explosion. The initial structure of the jet is ``top-hat'', meaning that the post-shock density, pressure, and velocity are constant with polar angle. The study of polarisation from structured jets is left for future work.

The jet propagates into a medium with a number density of $n=1$ cm$^{-3}$. The dynamics of strong shocks do not depend on the pressure (or temperature) of the surrounding medium, which is initialised in the simulation as $p=10^{-10} \rho c^2$. The simulation uses a 2D grid in spherical (polar) coordinates, with a radial and angular size of $(r_{\rm max}, \theta_{\rm max}) = (1.1\times 10^{19}$~cm, $\pi/2)$. The inner  boundary is located at $1.8 \times 10^{17}$~cm. The simulation runs for 150 years. A more detailed description of the initial conditions can be found in \citet{decolle12a}.

\subsection{Synchrotron emission}

The numerical simulations provide the post-shock energy density $e$, density $\rho$, and velocity $\mathbf{v}$ in each computational cell at different evolutionary phases.
To compute the synchrotron radiation, we post-process a large number of snapshots (one thousand) saved during the simulation. We assume that a fraction $\chi_e$ of the post-shock electrons is accelerated to relativistic speeds, creating a population of electrons with a density $n(\gamma_e) \propto \gamma_e^{-p}$ for $\gamma>\gamma_m$ and $n(\gamma_e) = 0$ for $\gamma<\gamma_m$, where $p$ is the power-law index of the population of non-thermal electrons accelerated by the shock, and $\gamma_m$ is the minimum Lorentz factor of the accelerated electrons, given by 
\begin{equation}
  \gamma_m(t) = \frac{p-2}{p-1} 
      \frac{\epsilon_e e_e}{\chi_e n_e m_e c^2}\;.
\end{equation}
The energy density of the accelerated electrons ($e_e$) and the post-shock magnetic field energy density $B^2/8 \pi$ (which determines the synchrotron emission intensity) are taken as a fraction $\epsilon_e$ and $\epsilon_B$ of the post-shock thermal energy respectively, i.e., $e_e=\epsilon_e e$ and $B^2/8\pi=\epsilon_B e$.

To compute the radiation received by an observer located at a distance $d_L$ and at an observing angle $\theta_{\rm obs}$ (measured with respect to the $z$-axis - being $z=r\cos\theta$), we remap the results of the 2D numerical simulation along the (azimuthal) $\phi$ direction.
Then, we compute the flux in each cell. 
We divide the range of observed times into $N_j$ bins, logarithmically spaced, with width $\Delta t_{\rm obs,j}$, and add each contribution to the corresponding time bin in the observing frame, by using the relation 
\begin{equation}
ct_{\rm obs}= (1+z) \left(ct - r \sin\theta \cos \phi \sin\theta_{\rm obs}- r \cos\theta \cos\theta_{\rm obs}\right)\;,
\label{eqtobs}
\end{equation}
where $r$ is the radial distance from the central engine, $\theta$ and $\phi$ the polar and azimuthal angles respectively, and $\theta_{\rm obs}$ is the observer angle, measured from the direction of propagation of the jet, and $z$ is the redshift.

At each evolutionary time $t$, we compute the differential flux $dF_{\nu,i}$ in each cell $i$, with $i=1,\dots,N_{\rm cells}$ being $N_{\rm cells}$ the total number of cells in each time frame as \citep{decolle12a}
\begin{equation}
    dF_\nu = \frac{1+z}{d_L^2}
    \frac{dV_i \Delta t_{\rm sim}}
{\Delta t_{\rm obs,j}}
\frac{j'_{\nu'}}
{\Gamma^2 (1-\hat{n}\cdot \vec{\beta})^2 }\;,
\end{equation}
where $dV_i$ is the volume of the cell $i$, $\Delta t_{\rm sim}$ is the time interval between two different snapshots of the numerical simulation, $\Gamma$ and $\vec{\beta}$ are the Lorentz factor and velocity of the fluid in the cell, $\hat{n}$ is the direction of the observer, and 
$j'_{\nu'}$ is the emitted energy per unit volume, solid angle, frequency and time. It is a function of the proper frequency $\nu'$, of the cooling frequency $\nu_c$, corresponding to the maximum Lorentz factor of the accelerated electrons (determined by cooling, i.e. $\nu_c\propto \gamma_c^2$) and of the characteristic electron frequency $\nu_m$ (corresponding to the minimum Lorentz factor of the initial population of accelerated electrons, i.e. $\nu_m\propto \gamma_m^2$).
In detail:
$j'_{\nu'}\propto {(\nu'/\nu'_m)^{1/3}}$  if $\nu'<\nu'_m<\nu'_c$, 
$j'_{\nu'}\propto {(\nu'/\nu'_m)^{1/3}}$  if $\nu'<\nu'_c<\nu'_m$, 
$j'_{\nu'}\propto {(\nu'/\nu'_m)^{(1-p)/2}}$ if $\nu'_m<\nu'<\nu'_c$, 
$j'_{\nu'}\propto {(\nu'/\nu'_c)^{-1/2}}$ if  $\nu'_c\nu'<\nu'_m$, $j'_{\nu'}\propto {(\nu'/\nu'_m)^{(1-p)/2}(\nu'/\nu'_c)^{-1/2}}$ if $\nu'>\nu'_m,\nu'_c$. 

The frequency in the proper frame is related to the one in the lab frame by the relation
\begin{equation}
    \nu' = (1+z) \Gamma (1-\hat{n}\cdot \vec{\beta}) \nu
\end{equation}
For a more detail description of the equations employed, we refer the interested reader to \citet{decolle12a}.

\subsection{Polarisation}

In this Section, we describe in detail the method employed to determine the Stokes parameters from our numerical simulations. Although our description is applied to 2D numerical simulations, it can be easily generalised to the three-dimensional case.

We compute the intensity $I$ and the Stokes parameters $Q$ and $U$ as \citep[see, e.g.,][]{Gill&Granot2020}
\begin{eqnarray}
I_\nu&=&\int dF_{\nu}(\sin \chi')^{\alpha+1}\;,
\label{poldegree1}
\\
Q_\nu&=&\int dF_{\nu}(\sin \chi')^{\alpha+1}\pi_\nu(\cos 2\tilde{\chi})\;,
\label{poldegree2}
\\
U_\nu&=&\int dF_{\nu}(\sin \chi')^{\alpha+1}\pi_\nu(\sin 2\tilde{\chi})\;.
\label{poldegree}
\end{eqnarray}
where $\tilde{\chi}$ is the position angle, and $\chi'$ is the angle between the magnetic field and the direction of the observer (both defined in the proper frame), i.e., 
\begin{eqnarray}
\cos \chi' = \hat{B}' \cdot \hat{n}'\;, \qquad
\sin \chi' = (1-\cos^2 \chi')^{1/2}\;.
\end{eqnarray}

As the simulation is axisymmetric, $U=0$ due to the symmetry of the $\sin$ function with respect to the $x$-axis\footnote{We checked that, effectively, the value of $U$ remains close to zero in our 2D simulations, with an error (small, i.e. such that $U\ll Q,I$) which depends on the number of samples used along the azimuthal angle $\phi$. This would be not true in three-dimensional, asymmetric simulations, or in 2D simulations in which the  magnetic field is taken as not symmetric with respect to the $x$-axis.}. Then, the integral value of the position angle, computed as $\chi = \frac{1}{2} \arctan (U/Q)$, depends on the sign of $Q$, with the polarisation direction forming an angle of 0$^\circ$ or 90$^\circ$ with respect to the reference direction $\hat{l}$ (defined below).

Furthermore, in equation \ref{poldegree}, $\pi_\nu$ is the local linear PD obtained by integrating the synchrotron emission over  a power-law distribution of relativistic electrons in each fluid element (i.e., in each cell in our case). Hereafter, we will indicate the ``local'' (i.e., defined for a single fluid element) PD as $\pi_\nu$, and the ``global'' PD (integrated over all the simulation volume) by $\Pi_\nu$.

\begin{figure}
    \centering
    \includegraphics[width=\columnwidth]{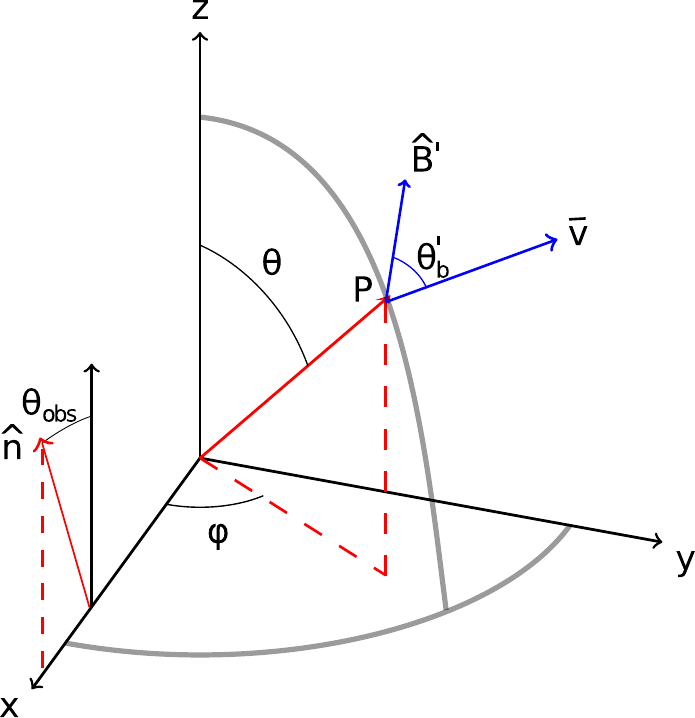}
    \caption{Diagram showing the coordinate system used in the calculation of the linear polarisation. The jet's main axis of propagation is aligned with the $z$ axis. $\hat{n}$ represents the direction of the observer, and it subtends an angle $\theta_{\rm obs}$ with respect to the $z$ axis. In each point $P$, the simulation provides the velocity $\hat{v}$, the density $\rho'$, and the thermal energy density $e'_{\rm th}$. The magnetic field $\hat{B}'$ subtends an angle $\theta_b'$ with respect to $\vec{v}$. }
    \label{fig1}
\end{figure}

Figure \ref{fig1} shows the geometry of the problem studied. The simulated jet is symmetric with respect to the $z$-axis. The observer is located at a polar angle $\theta_{\rm obs}$ with respect to the jet axis. The direction of the observer is then given (in the lab frame) by the vector  $\hat{n}=(\sin\theta_{\rm obs},0, \cos\theta_{\rm obs})$. In the proper frame, the observer's direction is \citep{lyutikov03}
\begin{eqnarray}
  {\hat{n}}'=\frac{
 {\hat{n}}+ \Gamma \vec{\beta} \left( \frac{\Gamma}{\Gamma+1} {\hat{n}}\cdot \vec{\beta} - 1
 \right) }{\Gamma (1-\hat{n}\cdot \vec{\beta})}\;,
\end{eqnarray}
where $\beta=v/c$.
We consider the synchrotron radiation coming from a region with volume $dV$ located at the position $P$ (see Figure \ref{fig1}). The electric field of a linearly polarised electromagnetic wave and the fluid magnetic field (both measured in the proper frame) are perpendicular with respect to each other, and related by $\hat{e}' = \hat{n}' \times \hat{B}'$ (with $\hat{e}'$ and $\hat{B}'$ both unit vectors). The electric field in the observer frame is \citep{lyutikov03}:
 \begin{eqnarray}
   \hat{e} &=&\frac{\hat{n}\times \vec{q'}}{\sqrt{q'^2-(\hat{n}\cdot \hat{q'})^2}} \;, \\
   \vec{q'} &=& \hat{B'}+ \hat{n} \times (\vec{\beta}\times \hat{B'})- \frac{\Gamma}{\Gamma+1}(\hat{B'}\cdot \vec{\beta}) \vec{\beta}   \;.
  \end{eqnarray}
The position angle $\tilde{\chi}$ (used in equation \ref{poldegree} to compute the Stokes parameters) corresponds to the angle between the polarisation vector and a given direction $\hat{l}$ in the plane of the sky, which (again, following \citealt{lyutikov03}) we take as the direction of the $y$ axis (which has the same direction in the lab and observer frame), i.e. $\hat{l}=(0,1,0)$.
Then, $\tilde{\chi}$  is determined by the relations
\begin{eqnarray}
\sin \tilde{\chi}&=&\mathbf{\hat{e}}\cdot \mathbf{\hat{l}}\;, \\
\cos \tilde{\chi}&=&\mathbf{\hat{e}}\cdot (\mathbf{\hat{n}} \times \mathbf{\hat{l}})\;.
\end{eqnarray}
The position angle $\tilde{\chi}$ is uniquely determined, given the velocity $\vec{v}$ of the parcel of fluid and the local magnetic field direction $\hat{B}$ (defined in the laboratory frame in these equations). 
In each computational cell, the velocity comes directly from the numerical simulations. 

In this paper, we consider several geometries for the magnetic field in the post-shock region.
As the ambient medium is at rest, the velocity of a parcel of fluid just behind the shock is perpendicular to the shock front itself. Furthermore, we assume that the post-shock magnetic field remains frozen in the fluid. Then, if we choose a direction for the magnetic field with respect to the local velocity once the parcel crosses the shock, it will conserve this direction as it is advected through the post-shock region. Thus, changes in the velocity of the fluid in the post-shock region will be associated with changes in the direction of the post-shock magnetic field.

\begin{figure}
    \centering
      \includegraphics[width=\columnwidth]{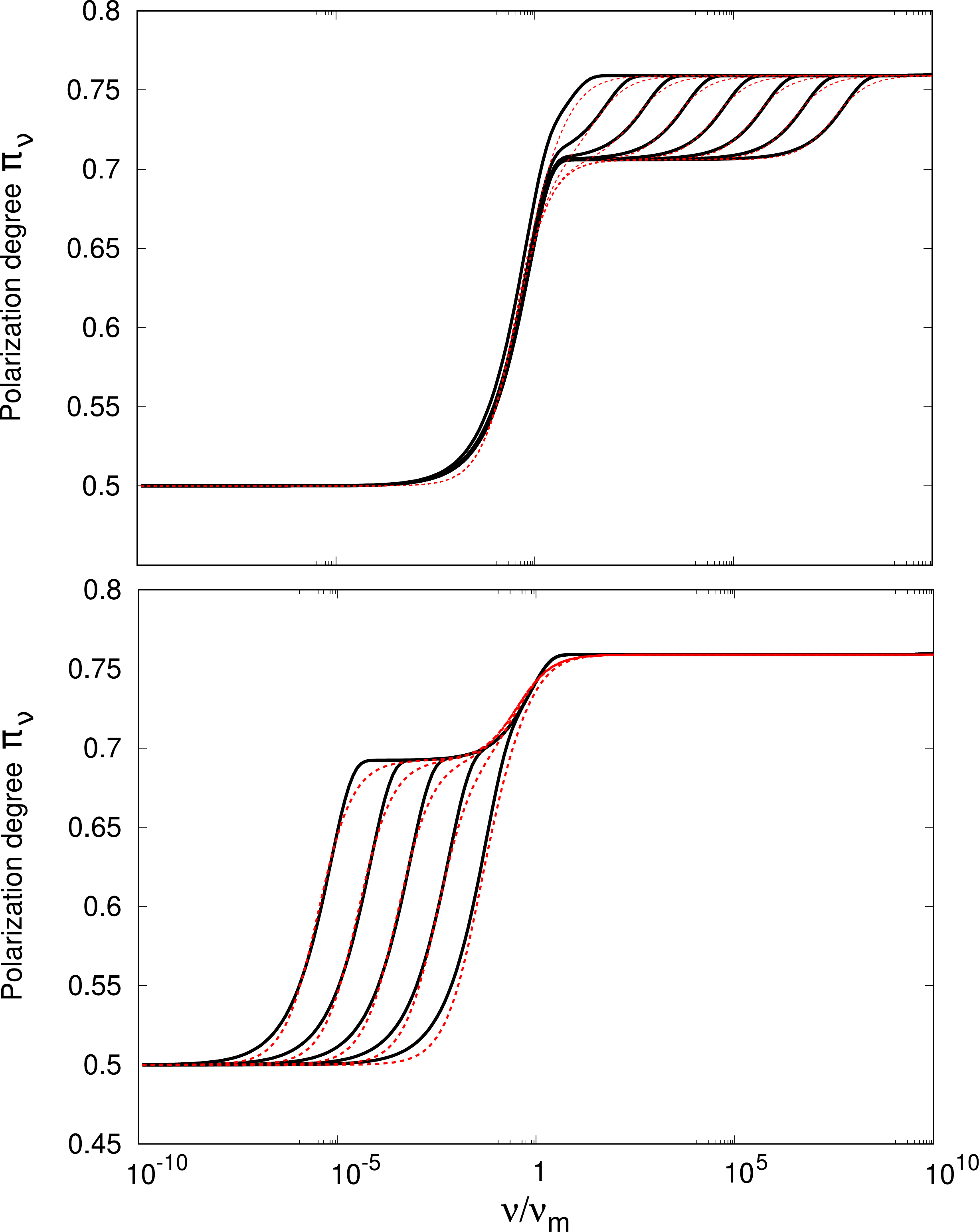}
    \caption{Local PD $\pi_\nu$ as a function of frequency. The upper panel shows $\pi_\nu$ in the slow cooling case, with different curves corresponding to different values of the cooling frequency, i.e., $\nu_c/\nu_m=10^8, 10^7, 10^6, 10^5, 10^4, 10^3, 10^2, 10^1$ (right to left curves). The bottom panel shows $\pi_\nu$ in the fast cooling case, with $\nu_c/\nu_m=10^{-1}, 10^{-2}, 10^{-3}, 10^{-4}, 10^{-5}$ (right to left curves). Black lines represent the results of the numerical calculation, while red lines show a simple fit that captures the general behaviour of the curves (see main text for more details).
    }
    \label{fig2}
\end{figure}

Given the surface $S$ tangent to the shock, in each cell we define a system of reference described by the following orthonormal basis: the unitary vector $\mathbf{\hat{n}_\parallel}$ perpendicular to $S$ and parallel to the velocity of the shocked material $v$ just behind the shock, and two orthonormal vectors $\mathbf{\hat{n}_{\theta}}$ and $\mathbf{\hat{n}_{\phi}}$ in the plane $S$, along the azimuthal and polar directions respectively. As a function of the velocity components 
\begin{eqnarray}
  \mathbf{v}=(v_r\cos\phi, v_r\sin\phi, v_z)\;, 
\end{eqnarray}
these vectors are given as
\begin{eqnarray}
\hat{i}_{\theta} &=& (v_z\cos\phi,v_z\sin\phi,-v_r)/v\;,\\
\hat{i}_\parallel &=& (v_r\cos\phi,v_r\sin\phi,v_z)/v\;,\\
\hat{i}_{\phi}&=& (-\sin\phi,\cos\phi,0)\;.
\end{eqnarray}
This orthonormal basis is then used to define a magnetic field $\vec{B}'$ in the proper frame as
 \begin{eqnarray}
   \hat{B}' = \cos \theta'_b \hat{i}_\parallel + \sin \theta'_b  \cos \phi' \hat{i}_{\theta} + \sin \theta'_b \sin \phi' \hat{i}_{\phi}\;,
\end{eqnarray}
where the angle $\theta'_b$ is the angle subtended by the vectors $\vec{v}$ and $\hat{B}'$ (see figure \ref{fig1}), and $\phi'$ defines the direction of the magnetic field in the plane parallel to the shock front\footnote{As $\hat{i}_\parallel$ is parallel to the velocity, the proper frame unit vectors $\hat{i}'_\parallel, \hat{i}'_\theta, \hat{i}'_\phi$  are identical to the lab frame unit vectors $\hat{i}_\parallel, \hat{i}_\theta, \hat{i}_\phi$.}. 
We consider different directions for the magnetic field. In particular, a  magnetic field parallel to the velocity $\hat{B}=\hat{n}_\parallel$ (corresponding to $\theta'_b=0$), a magnetic field perpendicular to the velocity (i.e., parallel to the shock plane) and random (corresponding to $\theta'_b=\pi/2$),  defined as $\hat{B}' = \cos \phi' \hat{i}_{\theta} + \sin \phi' \hat{i}_{\phi}$, and different combinations of radial and tangential magnetic fields. The fraction of parallel over perpendicular component is defined as 
 \begin{eqnarray}
   \eta = \frac{B'_\parallel}{B'_\perp} = \frac{\cos \theta'_b}{\sin \theta'_b}\;.
\end{eqnarray}

When the magnetic field is tangled in the post-shock region at a scale smaller than the size of a computational cell (as we assume in this paper), the linear polarization is computed by averaging over the different directions of the local magnetic field within each cell. In practice, we write the magnetic field in the shock plane as $\hat{B}'=\cos \phi' \hat{i_\theta}+\sin \phi' \hat{i_\phi}$, and we sample randomly the angle $\phi'$ 10 times in each cell.

\begin{figure*} 
\includegraphics[width=1.5\columnwidth]{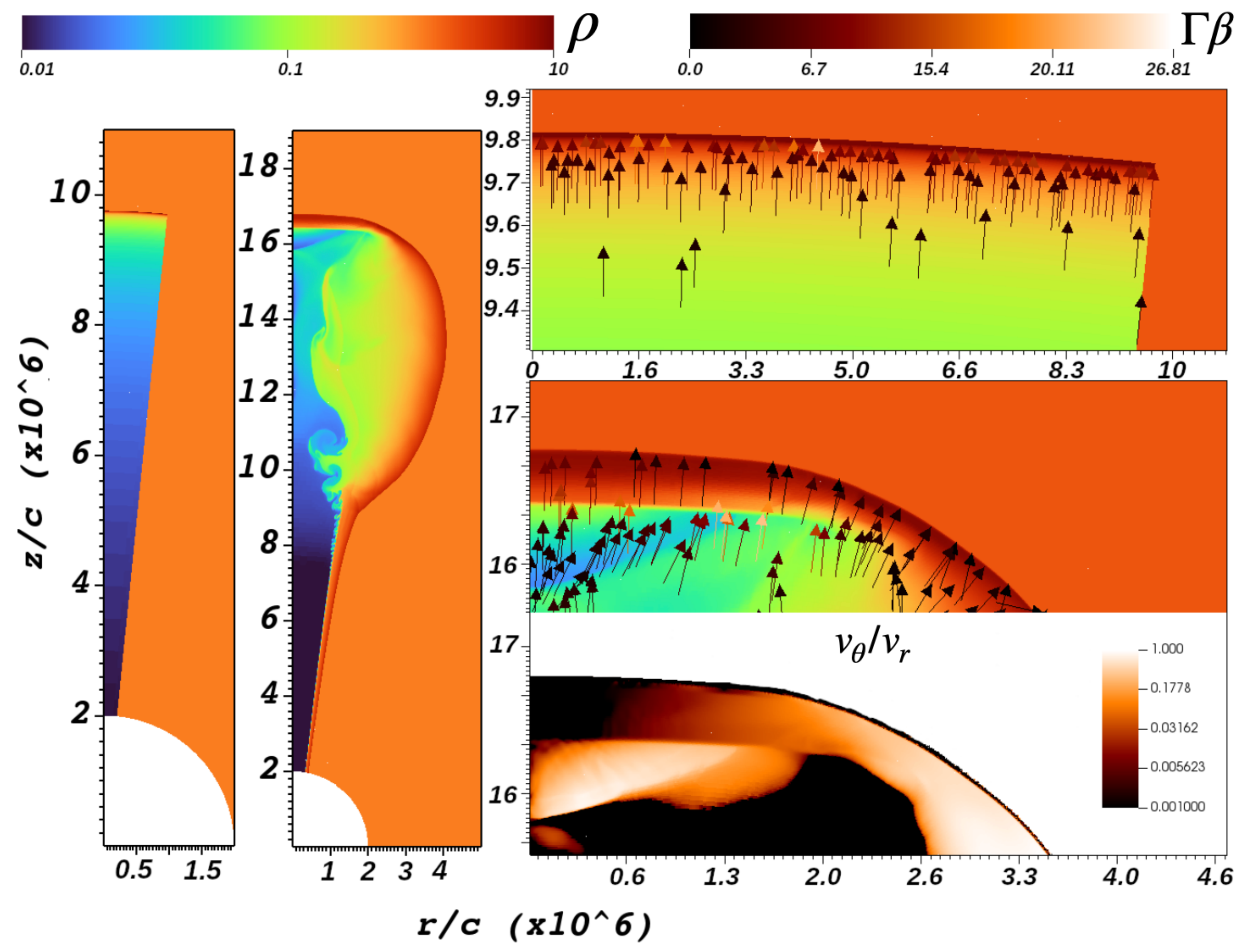}
    \caption{Density and velocity ($\Gamma \beta$) maps. The left panels show the density maps at $t=9.7 \times 10^6$~s (i.e. the initial condition of the simulation) and at $t=1.6 \times 10^7$~s,  showing the lateral expansion of the jet. The right panels show the velocity map $\Gamma\beta$ and the velocity field at the same evolutionary times (top and middle panels), and the ratio of tangential over radial velocity (bottom).
    }
    \label{fig3}
\end{figure*}

The local PD $\pi_\nu$ depends on the particular spectral range considered, i.e. on the value of the frequency $\nu$ with respect to the cooling frequency $\nu_c$ and the characteristic electron frequency $\nu_m$.
Far from the characteristic frequencies,  we have
$\pi=(\alpha+1)/(\alpha+5/3)$, where  $\alpha$ is the spectral index of the synchrotron spectrum. Its value depends on the particular spectral range considered \citep[see, e.g.,][]{Gill&Granot2020}.
When $\nu\ll\nu_m,\nu_c$, we have $\alpha=-1/3$ and $\pi_1=1/2$. When $\nu\gg\nu_m,\nu_c$, we have $\alpha=p/2$ and $\pi_3=(p+2)/(p+10/3)$. When $\nu_m\lesssim\nu\lesssim\nu_c$ (in the slow cooling regime), we get $\alpha=(p-1)/2$ and $\pi_{2,s}=(p+1)/(p+7/3)$
Finally, when $\nu_c<\nu<\nu_m$ (in the fast cooling regime) we get $\alpha=1/2$ and $\pi_{2,f}=9/13$.

These analytical scaling for $\Pi_\nu$ have been usually employed to compute the PD.
\citet{Birenbaum&Bromberg2021} showed that the PD changes smoothly (over several orders of magnitude) from one characteristic value ($\pi_1, \pi_{s2}, \pi_{f2}, \pi_3$) to the other. As this can have potentially an impact on the calculation of the (integrated) PD, we computed the dependence on the local PD $\pi_\nu$ as a function of frequency, given as the ratio
\begin{eqnarray}
    \pi_\nu=\frac{\int G(\nu/\nu_s) N(\gamma) d\gamma}{\int F(\nu/\nu_s) N(\gamma) d\gamma}\;,
    \label{eqfnu}
\end{eqnarray}
where $G(x)=x K_{2/3}(x)$ and $F(x)=x \int_x^\infty K_{5/3}(t) dt$, being $K_m$ the modified Bessel function of order $m$ \citep[see, e.g.,][]{Rybicki&Lightman1979}.
As electrons cool due to synchrotron radiation, their energy evolves as $dE/dt = m_e c^2 d\gamma_e/dt = - \sigma_T c B^ 2 \gamma^2/(6\pi)$, being $m_e, c, \sigma_T$ the electron mass, the speed of light and the Thomson cross-section respectively. Then, the Lorentz factor of each electron changes with time as
\begin{eqnarray}
  \gamma_e = \frac{\gamma_{e,0}}{1+k \gamma_{e,0} t}\;,
\end{eqnarray}
where $\gamma_{e,0}$ is the initial Lorentz factor of the electron, and $k= 6\pi m_e c/(\sigma_T B^2)$.
To compute $\pi_\nu$, we integrate the emission of electrons with a Lorentz factor $\gamma_e$ over the electron population and over time. When electrons cool for a short time, i.e. $\gamma_c=1/kt\gg\gamma_m$, implying $\nu_c\gg \nu_m$, we get slow cooling, while fast cooling is obtained by leaving electrons cool much longer, i.e. $\gamma_c = 1/kt\lesssim\gamma_m$, corresponding to $\nu_c\ll \nu_m$.

Figure \ref{fig2} shows the PD computed for slow and fast cooling (top and bottom panels respectively). The different curves correspond to different values of the ratio $\nu_c/\nu_m$.
Consistently with \citet{Birenbaum&Bromberg2021}, the transition between the different constant regions is smooth, and extends over several orders of magnitude in frequency.
As integrating equation \ref{eqfnu} for each computational cell would be computationally expensive, we instead used the following simple fitting functions (dotted red lines in figure \ref{fig2}), which capture the general behavior of $\pi_\nu$:
\begin{eqnarray}
   \pi = \pi_1 + \frac{(\pi_2-\pi_1)\nu/\nu_m}{\nu/\nu_m+1/3} + \frac{(\pi_3-\pi_2)\nu/\nu_c}{\nu/\nu_c+1/3}
   \label{eqfit1}
\end{eqnarray}
for slow cooling ($\nu_c>\nu_m$), and
\begin{eqnarray}
   \pi = \pi_1 + \frac{(\pi_2-\pi_1)\nu/\nu_c}{\nu/\nu_c+1/3} + \frac{(\pi_3-\pi_2)\nu/\nu_m}{\nu/\nu_m+1/3}
   \label{eqfit2}
\end{eqnarray}
for fast cooling ($\nu_m>\nu_c$). It is easy to verify that in the case $\nu_m=\nu_c$ the two solutions converge to the same solution.

\section{Results}

In this section, we show the time evolution of the PD computed by post-processing the results of hydrodynamic simulations (see section \ref{sec:methods}), for both on-axis and off-axis observers. We present results obtained by considering a top-hat jet decelerating through a uniform medium. The extension to jets propagating through a windy medium (i.e. with a density profile  $\rho\propto r^{-2}$) and to structured jets will be considered in future work.

Figure \ref{fig3} shows density and velocity maps ($\Gamma\beta$) at $t=112.27$ days (i.e., at the beginning of the simulations, the first panel from the left) and at a later evolutionary stage ($t=185.18$ days, second panel from the left). The initial density and velocity are defined by considering the \citet{blandford76} self-similar solution. As the shock wave decelerates, it expands laterally, so that also assuming an initial top-hat jet, after a time corresponding to when $\Gamma_{\rm sh}(t) \lesssim 1/\theta_j$, being $\Gamma_{\rm sh}$ the shock Lorentz factor, the jet acquires a lateral structure.
The right panels of figure \ref{fig3} show maps of the velocity 4-vector (at the same times as the left panels) and the velocity field. While initially all velocities are radial (see top panel), lateral expansion twists the velocity field lines and creates a large tangential velocity component at the edge of the jet (see the bottom panel of the figure). This is the main difference between numerical simulations and analytical models, in which typically the velocities are taken as radial at all times. Assuming that the magnetic field is frozen in the fluid, the tangling of the velocity field implies that also the magnetic field direction will change in the post-shock region. We consider this magnetic field direction when calculating the Stokes parameters (see equation \ref{poldegree}).

\begin{figure}
    \includegraphics[width=\columnwidth]{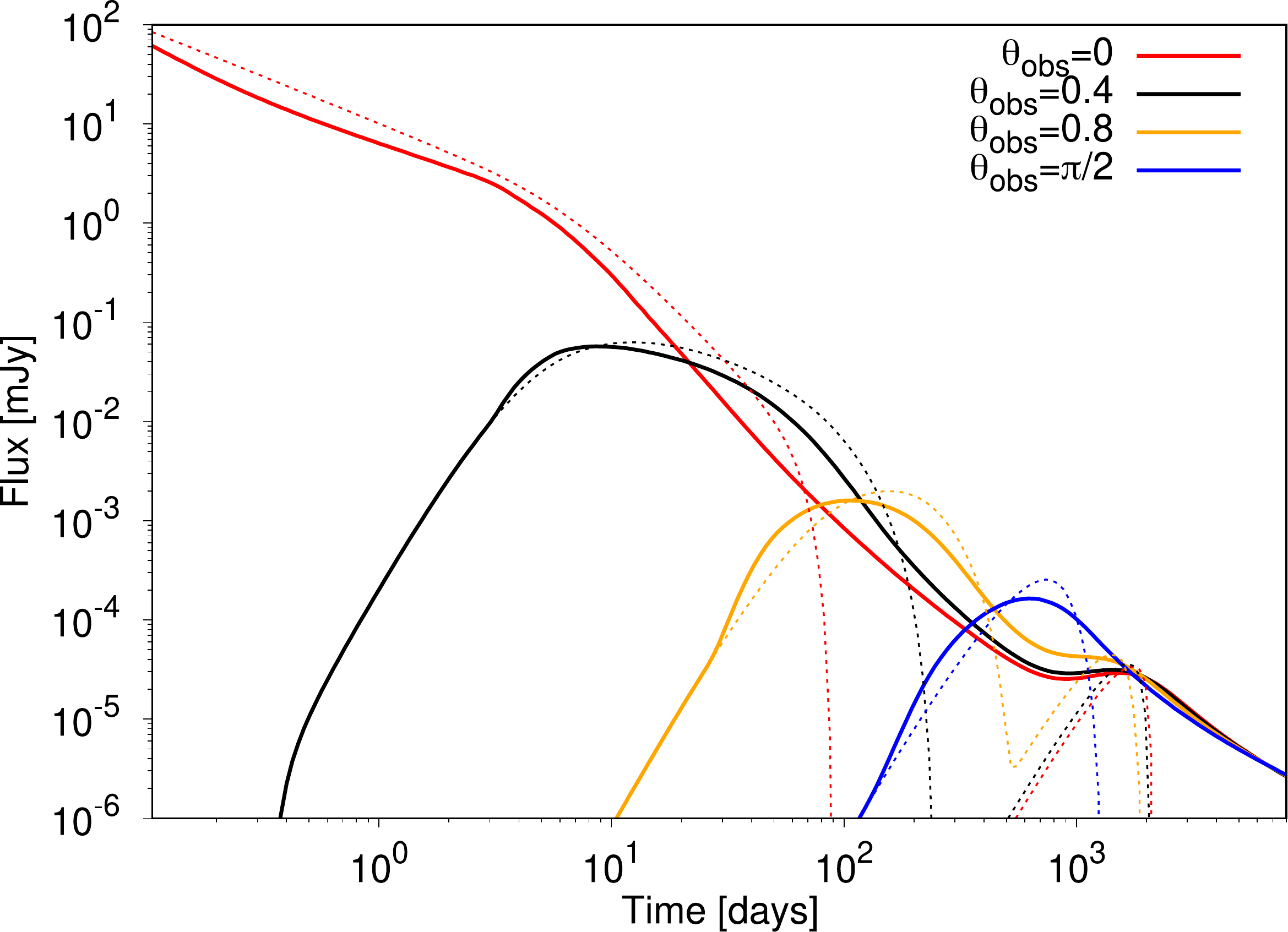}
    \caption{Optical ($\nu_{\rm obs} = 4.556 \times 10^{14}$ Hz) light curves at different observing angles ($\theta_{\rm obs}=0,0.2,0.4,0.8,\pi/2$ rad).
    The source has been placed at a distance $d = 40$ Mpc. The microphysical parameters are $\epsilon_e = 0.1$, $\epsilon_B = 10^{-3}$ and $p=2.2$. The dotted lines in the figure represent the light curves computed by assuming a one-dimensional, analytical solution (i.e. neglecting lateral expansion). The decelerating relativistic shell has an isotropic energy $E_{\rm iso} = 10^{53}$ erg.}
    \label{fig4}
\end{figure}

As a reference to interpret the time evolution of the polarisation, we show in figure \ref{fig4} the optical afterglow light curve (at a frequency $\nu=4.5\times 10^{14}$ Hz) for different observer angles ($\theta_{\rm obs} =0, 0.4, 0.8, \pi/2$). The on-axis light curve shows a jet break at $\sim 3$ days, corresponding to the time when the edge of the jet becomes visible to the observer, due to the deceleration of the jet itself. 
aaaaaaaaaaa\citep[e.g., ][]{rhoads97, sari99a, kumar00}
At larger observing angles, light curves peak at later times, as they enter the field of view of the off-axis observer. 
The increase in the flux at $\sim$ 1000 days is due to the appearance of the counter-jet. At late times ($\gtrsim$ 1000 days), the jet becomes sub-relativistic and the light curves become independent of the observing angle. The figure also shows analytical light curves computed by considering the deceleration of a jet wedge, described by the \citet{blandford76} self-similar solution during the relativistic phase. The analytical light curves are similar to the numerical light curves. The main differences are present after the jet break, when the analytical curves (which do not include lateral expansion) overestimate the flux by a factor of a few. At late times, as the \citet{blandford76} self-similar solution is valid only in the ultra-relativistic regime (i.e. when the shock Lorentz factor is $\gg 1$), the flux computed from the analytical solution goes to zero.

\begin{figure}
\includegraphics[width=\columnwidth]{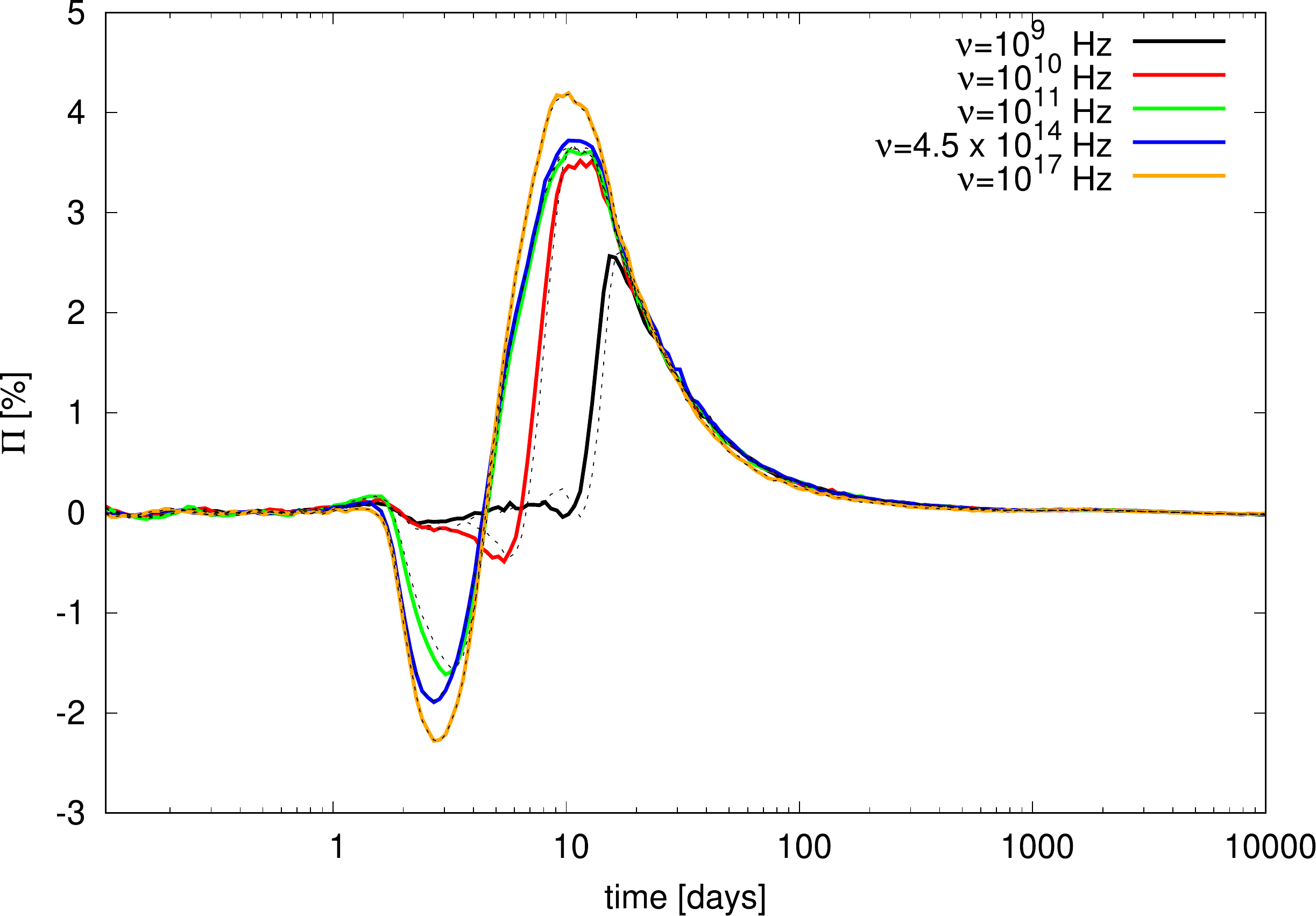}
\caption{Time evolution of the PD for a nearly on-axis observer (located at $\theta_{\rm obs}=0.05$ rad), for different frequencies: 1 GHz, 10 GHz (radio), 100 GHz (microwave), $4.5\times 10^{14}$ Hz (optical) and $10^{17}$ Hz (X-rays). The jet characteristics and the parameters used to compute the synchrotron emission are the same as those of figure \ref{fig4}.
The magnetic field is tangential to the shock plane, i.e., perpendicular to the velocity vector. The polarisation is computed by considering a fit for the local polarisation (equations \ref{eqfit1} and \ref{eqfit2}), while the dotted lines correspond to $\pi_\nu$ constant in each spectral slope segment.}
\label{fig5} 
\end{figure}

Figure \ref{fig5} shows the PD for a tangential random magnetic field, for different frequencies, for an observer located at $\theta_{\rm obs} = 0.05$~rad, i.e. inside the initial jet opening angle $\theta_j=0.2$~rad\footnote{
The polarisation degree is defined as $\Pi_\nu = \sqrt{Q^2+U^2}/I$. As the Stokes parameter $U=0$ in our case, we use instead the alternative definition $\Pi_\nu = Q/I$, allowing it to take negative values.}. The light curves at $\nu>10^{11}$~Hz present a similar behavior (the curves at $\nu=10^{12}$ Hz and $\nu=10^{13}$ Hz, not shown in the figure, are identical to the optical curve), with a minimum at $\sim 3$ days (with $\Pi\sim -1.5/-2.2$) followed by a steep increase and a maximum (corresponding to $\Pi\simeq 3.5\%$) at $\sim 10$ days. 
The transition from negative values of $\Pi$ (i.e. $Q<0$) to positive values corresponds to a rotation by 90$^\circ$ of the position angle \citep{sari99, ghisellini99}.

The polarisation in radio frequencies ($\nu=10^9-10^{10}$~Hz) extends over a smaller range both in $\Pi$ and time. The curves peak at later time, as radio frequencies correspond to $\nu < \nu_m <\nu_c$ while optical and X-rays to $\nu_m <\nu < \nu_c$ for the jet and microphysical parameters used in our simulations. All frequencies present a similar late decay in time, $\propto t^{-1.2-1.3}$. After 100 days, the PD vanishes in all models. The emission from the counter-jet (at $\sim 2000$ days) has a negligible PD. 
Figure \ref{fig5} also shows a comparison between the PD computed using constant values of the local polarisation $\pi_\nu$ (dashed lines), and using the smoothed $\Pi_\nu$ (full lines, see equations \ref{eqfit1} and \ref{eqfit2}). The two cases are nearly indistinguishable.

\begin{figure}
\includegraphics[width=\columnwidth]{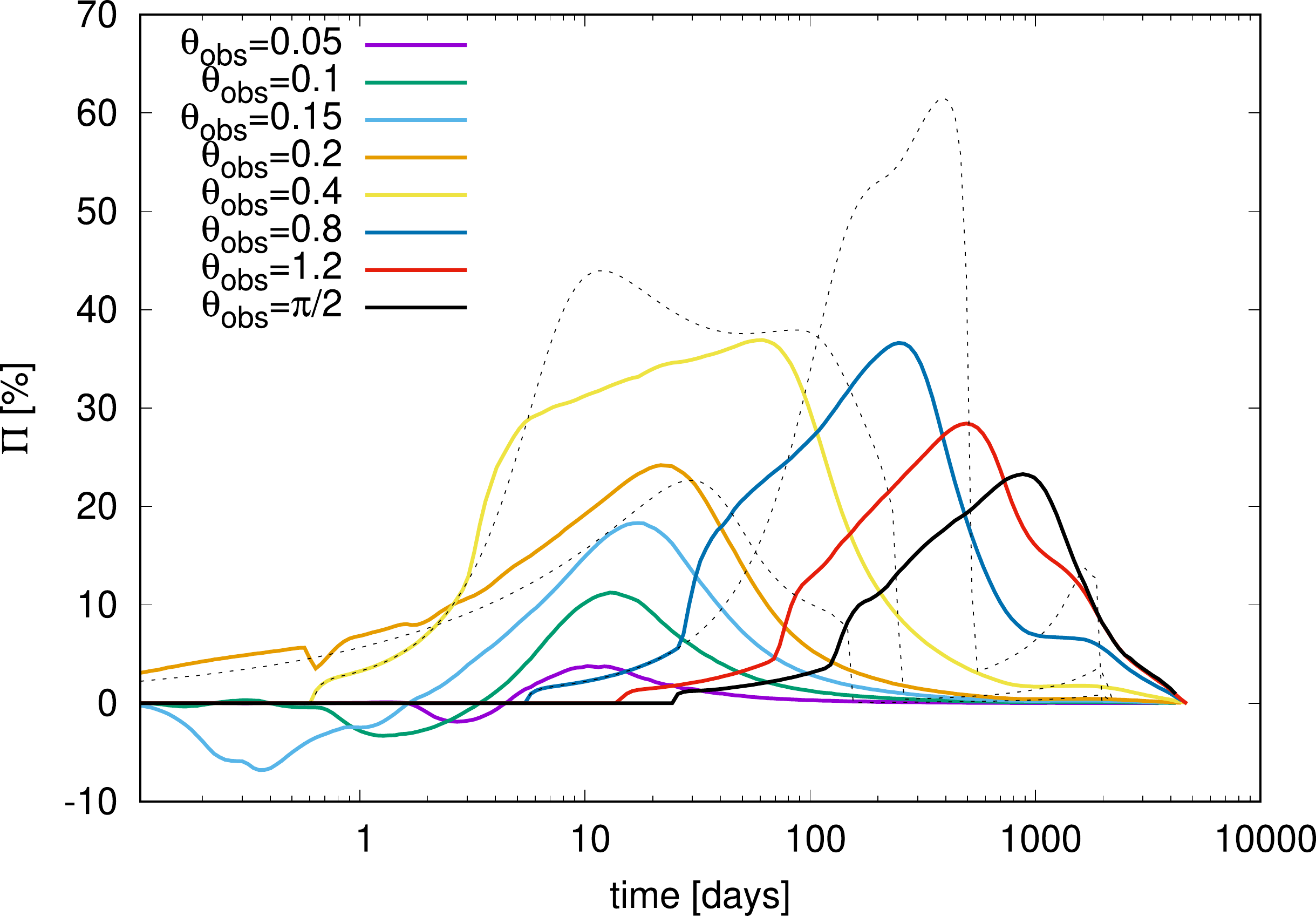}
   \caption{Time evolution of the PD for observers located on-axis ($\theta_{\rm obs}= 0.05, 0.1, 0.15, 0.2$ rad) and off-axis ($\theta_{\rm obs}= 0.4, 0.8, 1.2, \pi/2$ rad), for a magnetic field tangential to the shock front. The dotted lines show the values of $\Pi_\nu$ computed by considering the emission from a self-similar, decelerating wedge, with the same energy, opening angle, and ambient density employed in the hydrodynamical simulation, at observing angles $\theta_{\rm obs}= 0.2, 0.4, 0.8$ rad.}
    \label{fig6}
\end{figure}

Figure \ref{fig6} shows the optical afterglow polarisation curves, considering the same jet parameters as figures \ref{fig4} and \ref{fig5}, for different on-axis ($\theta_{\rm obs}<\theta_j=0.2$ rad) and off-axis angles ($\theta_{\rm obs}>\theta_j=0.2$ rad), and for a random magnetic field in the shock plane. For an observer located completely on-axis ($\theta_{\rm obs}=0$ rad), $\Pi_\nu=0$ as the simulation and the magnetic field geometry are both axisymmetric. For observers located on-axis, $\Pi_\nu$ presents always a negative minimum (when, i.e., $Q<0$), and a maximum at times larger than the jet break time. The time corresponding to the minimum/maximum drops/increases for increasing observer angles, arriving at $\Pi\sim 25\%$ for an observer located at the edge of the jet.

A larger PD is obtained for off-axis observers. In this case, the curve is always positive, and the peak in $\Pi_\nu$ moves to larger times (again, larger than the jet break), arriving at a maximum between $\theta_{\rm obs}=0.4$~rad and $\theta_{\rm obs}=0.8$~rad (with $\Pi\sim 35\%$), then decreasing to $\sim 25\%$ and $\sim 20\%$ at $\theta_{\rm obs}=0.8$~rad and $\theta_{\rm obs}=\pi/2$~rad. The counter-jet produces a small change in the value of $\Pi$ (by about $\sim 5\%)$ at $t_{\rm obs}\gtrsim 1000$~days. Then, it can potentially be detected in radio for very close off-axis GRBs.
 
The effect of lateral expansion can be understood by comparing the polarisation computed from the numerical simulations (full lines of figure \ref{fig6}), in which the lateral expansion is a direct result of the evolution of the system, with the analytical self-similar solution (dotted lines), corresponding to the deceleration of a relativistic wedge (in which the lateral expansion is absent). Lateral expansion affects polarisation in different ways. At $\theta_{\rm obs}=0.4$~rad, $\Pi_\nu$ computed from the simulation presents a single peak while two peaks are produced in the analytical model. Finally, in analytical models, the peak in $\Pi_\nu$ increases with the observer angle at $\theta_{\rm obs}>0.4$~rad, while in the numerical simulations, the maximum value of $\Pi$ drops with the observer angle. 

\begin{figure}
\includegraphics[width=\columnwidth]{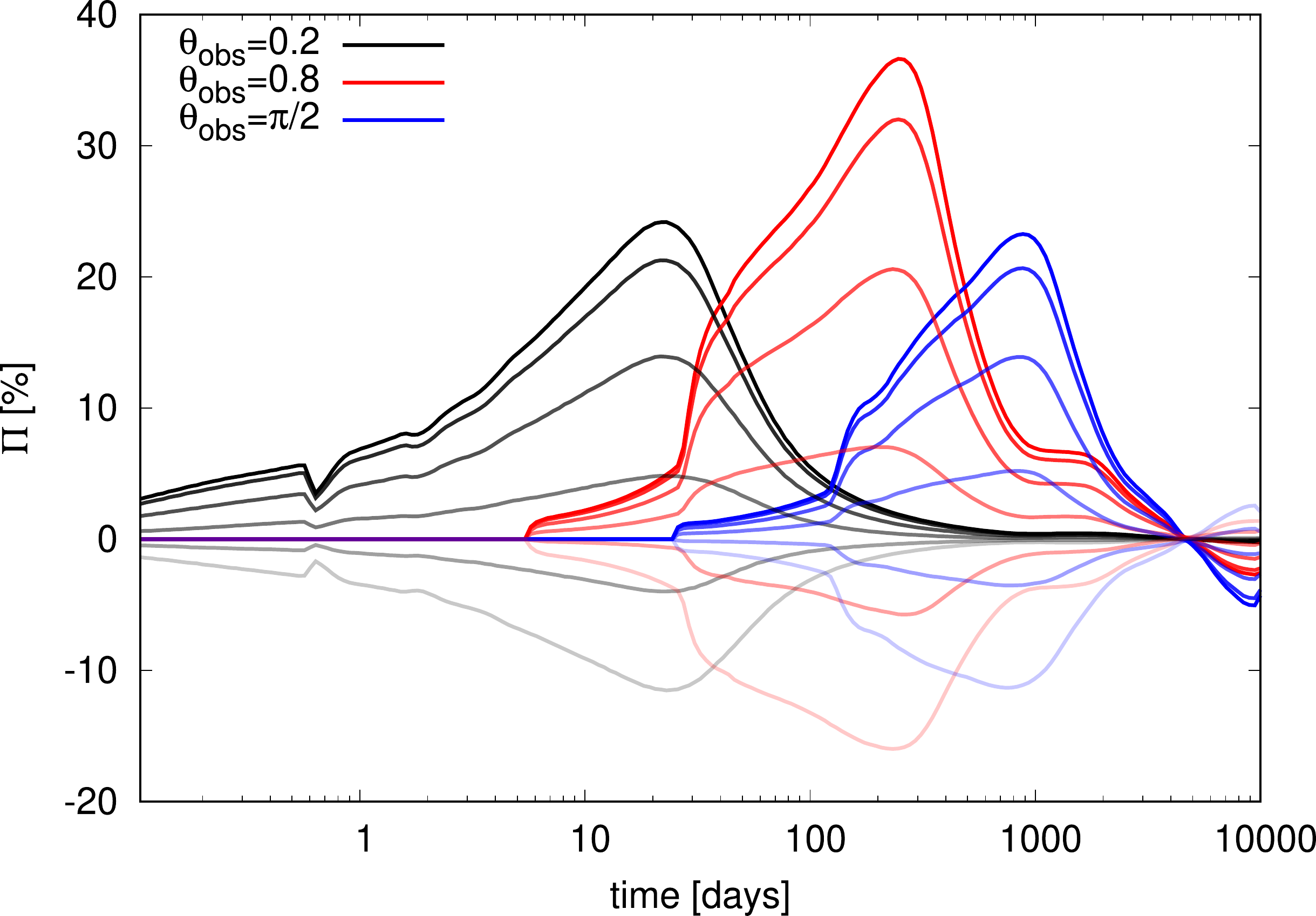}   \caption{Time evolution of the PD for an observer located on the jet edge ($\theta_{\rm obs}= 0.2$ rad) and off-axis ($\theta_{\rm obs}= 0.4, \pi/2$ rad), computed by considering a magnetic field perpendicular to the velocity (i.e. parallel to the shock plane) $B_\perp$ plus a parallel  component $B_\parallel$, for different ratios of the parallel vs perpendicular component (measured in the proper frame). From top to bottom curves, $B_\parallel/B_\perp = (0, 0.2,  0.4, 0.6, 0.8, 1)$.}
    \label{fig7}
\end{figure}

Figure \ref{fig7} shows the off-axis optical afterglow polarisation curve for a random magnetic field (parallel to the shock plane) plus a parallel component (both defined in the proper frame). The parallel component goes from 0\% to 100\% with respect to the tangential (randomly oriented) component. A parallel component produces a PD opposed in sign with respect to the perpendicular component. As mentioned in section 2, we compute the polarization sampling (randomly) the angle $\phi'$ 10 times. The small fluctuation visible at $\sim 1$ day in figure \ref{fig7} is a numerical artefact due to the low number of  angles $\phi'$ employed. Thus, increasing the parallel component leads to a decrease of PD. 
As shown by several authors \citep[e.g.,][]{gruzinov99a, sari99,granot03}, the local PD is $\propto (B_\parallel^2-B_{\perp}^2/2)$.  
Consistently with the results shown in figure \ref{fig7}, the curve corresponding to a null PD is obtained in the case of $B'_\parallel = B'_\perp/\sqrt{2}\simeq 0.7B'_\perp$, i.e. in the case of a completely isotropized magnetic field.

\section{Discussion}

\subsection{Polarisation Degree}

In this paper, we computed the linear polarisation associated with the afterglow emission of GRBs by using numerical simulations of a decelerating jet.
We computed the polarisation at several frequencies by considering different magnetic field geometries and observing angles. In this section, we discuss the main results of this work.

Several authors have theoretically studied  GRB polarisation during the afterglow phase. Although analytical methods typically allow us to understand the general behaviour of a system in a computationally inexpensive way, simulations are needed to obtain the detailed evolution of the system. Figure \ref{fig3} shows that once the GRB wedge starts decelerating, the lateral expansion deforms the shock structure, creating a large velocity component along the polar direction. The dynamics of the system are not correctly captured in calculations where the lateral expansion is not considered.

The time evolution of the PD computed from our simulations for an on-axis observer (see figure \ref{fig5}) has behaviour consistent with those obtained in previous semi-analytical models computing polarisation in several bands \citep{Birenbaum&Bromberg2021, Shimoda&Toma2021}. \citet{rossi04} modelled the lateral expansion by considering different expansion velocities, from a fraction of the sound speed to the relativistic sound speed. In their calculation, they obtained a drop in the polarisation $\gtrsim 50\%$ when considering lateral expansion, while in our case the polarisation $\Pi$ is of the same order as in the calculations by  \citet{Birenbaum&Bromberg2021, Shimoda&Toma2021}. 

\citet{Birenbaum&Bromberg2021} noticed that the local PD changes smoothly between the theoretically expected values valid in each frequency regime (see our figure \ref{fig2} and figure \ref{fig4} of \citealt{Birenbaum&Bromberg2021}). They claimed that polarisation in optical and microwave bands could be very different in the two cases. Surprisingly, our calculations show that this is not the case, and the differences in $\Pi$ in the two cases are minimal (see figure \ref{fig5}), at least for the set of parameters used in this paper. This is due to two reasons. First, the jump in the local PD is $\Delta\Pi = 4/(3(p+10/3)(p+7/3))\sim 0.05$ when crossing $\nu_c$ (for slow cooling). Thus, the maximum change in polarisation expected by using the two methods, in frequencies close to $\nu_m<\nu<\nu_c$, is $\Delta \Pi\sim0.05/0.7\lesssim 10\%$. Second, in our simulation the emission is not coming from a single parcel but is the sum of the contribution of volume elements localised at different times, positions, and angles (see equation \ref{eqtobs}) and travelling with different fluid velocities (e.g., with smaller radial velocity if more off-axis or at a later time). Thus, the proper frequency corresponding to a single observer frequency will be different in each fluid element, as well as the local PD.   

While the on-axis linear polarisation computed by employing numerical simulations is qualitatively similar to the one obtained by analytical methods, the off-axis polarisation differs in several ways.
The most notable difference is the presence of a single peak in the value of $\Pi$ computed at $\theta_{\rm obs}=0.4$ rad in our case, instead of two peaks obtained in analytical models \citep{granot03,rossi04}.
The lateral expansion affects the calculations for two reasons: 1) the jet plasma moves towards larger polar angles; 2) part of the radial velocity becomes tangential when expanding laterally. Then, the flux for an observer located on-axis is smaller (after the jet break) in the simulations with respect to the analytical models, while it increases for an observer located off-axis.
In the analytical model, the two peaks correspond to the time when the edge of the jet and the core of the jet become visible. Being the core more energetic, it produces a peak in the linear PD. In the simulations, on the other hand, the edge of the jet expands gradually before entering in the line of sight of off-axis observers. As a result, instead of a peak we get a slow increase in the PD. The case of a top-hat jet with the lateral expansion is then somehow similar to the case of a structured jet (in which $\Pi$ only presents a single peak, see \citealt{rossi04}). At later times, some of the energy located in the jet core also moves toward larger polar angles. Thus, the peak in polarisation drops with angle (instead of increasing as in analytical models). Analytical model overestimates the off-axis linear PD with respect to numerical simulations.
This illustrates the need to properly resolve the dynamics of the system to obtain a precise estimation of the radiative properties.

\subsection{Comparison with observations}

\begin{figure}
\includegraphics[width=0.9\columnwidth]{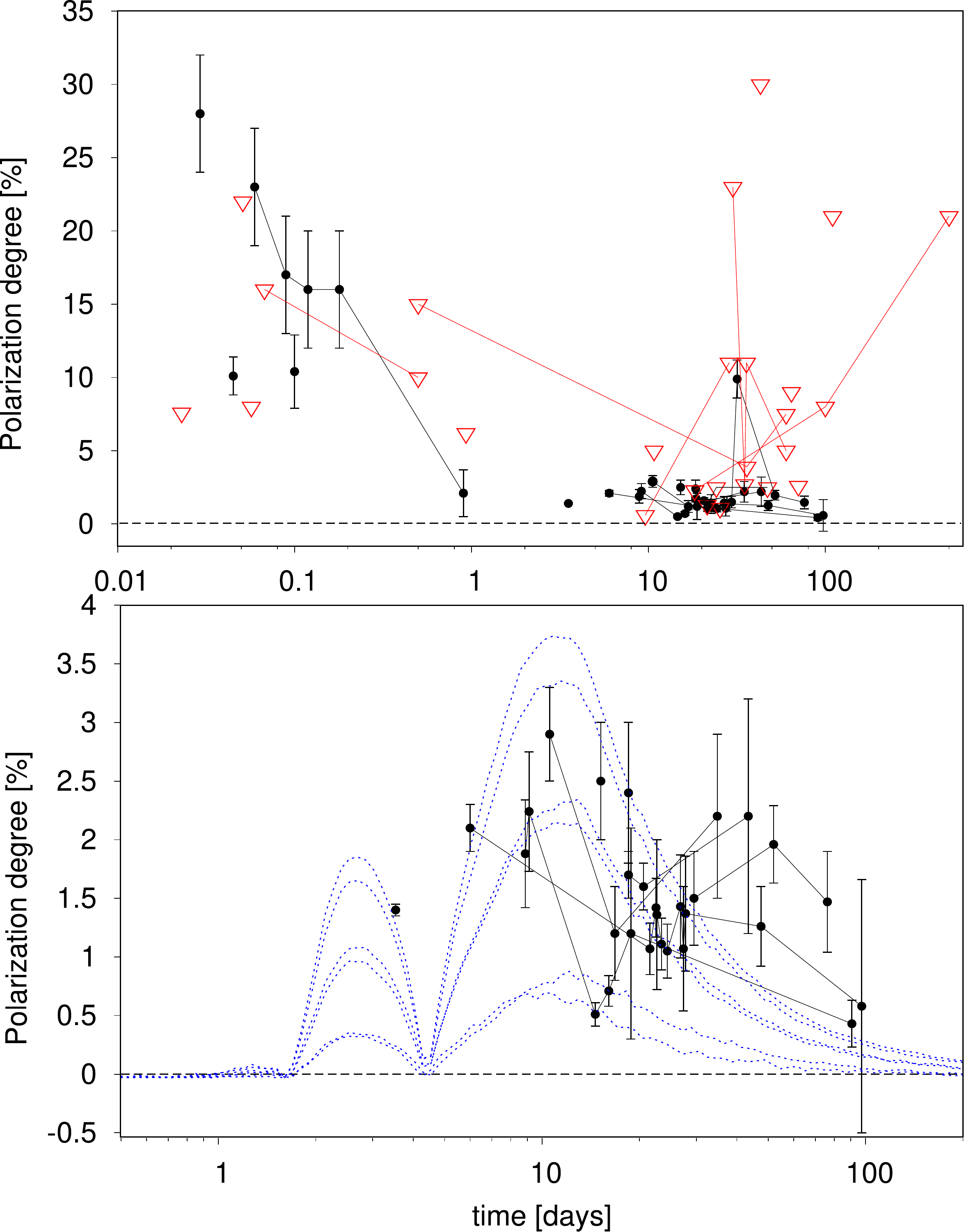}
\caption{\emph{Top panel}: Observations of linear polarisation in the afterglow emission of GRBs. Red points correspond to upper limits, while points connected by segments correspond to the same GRB. The following GRBs are included in the figure: GRB 990510, 990712, 010222, 020405, 020813, 020104, 030328, 080928, 090102, 091208B, 110205A, 120308A, 131030A \citep[see][and references therein]{Covino&Gotz16} 170817A \citep{Corsi18}, 171205A \citep{Urata19}, 190114C \citep{Laskar19,jordana-mitjans20}, 191221B \citep{Buckley21}.
\emph{Bottom panel}: a comparison between observations and the results of the numerical calculations, for $\theta_{\rm obs}=0.05$~rad, and $B_\parallel/B_\perp=0, 1, 0.2,0.8, 0.4,0.6$ (from the larger to the smaller blue dotted line)
}
    \label{fig8}.
\end{figure}

Linear polarisation has been detected in several GRBs. In figure \ref{fig8} (top panel) we present a sample of polarisation measurements obtained during the afterglow phase. In the figure, black points connected by lines indicate polarisation detection obtained for the same GRB, while isolated points represent cases where a single measurement is available. Red points and lines correspond to upper limits. Before $\sim$~1 days, the polarisation is relatively large ($\gtrsim 10\%$), while the values of linear polarisation detected at later times are much smaller. The presence of linear polarisation before $\lesssim$ 1 day is typically attributed to a reverse shock, where the jet head is still energised from new, fresh material, while late emission is associated with the forward shock. 

The bottom panel of figure \ref{fig8} presents a comparison between observations of polarisation due to the forward shock and the numerical model considered in this paper. In the figure, the dotted lines show the PD (defined here as $|Q|/I$) for optical frequencies (where most GRB polarisations have been detected). The same parameters as figure \ref{fig7} are used except for the observer angle, which is $\theta_{\rm obs}=0.05$~rad.
The intensity colour scale in the figure corresponds to different anisotropy factors (as in figure \ref{fig7}).

Qualitatively, the peak in polarisation restricts the anisotropy factor of the magnetic field to the range $B_\parallel/B_\perp \approx 0.4-1$. Nevertheless, we notice that the light curve and polarisation depend on (at least) six parameters: density of the CSM, jet energy, $\theta_{\rm obs}$, $\epsilon_e$, $\epsilon_B$, and $p$, in addition to the jet structure and magnetic field orientation. All of these parameters are fixed in the calculation presented in the figure, with values somehow representative of the GRB population. However, individual GRBs will in general have different values of these parameters.
For instance, the presence of polarisation at late times (i.e., with a peak at $\sim$50 days) could be explained by assuming that those specific jets have larger isotropic energy or lower density. The observer time scales as \citet{granot2012, vaneerten&macfadyen2012}
\begin{equation}
\frac{t'}{t}=\left( \frac{E'/E}{n'/n}\right)^{1/3},
\end{equation}
where $t$, $E$, $n$ are the values used in the simulation, and $t'$, $E'$, $n'$ are the rescaled values. To move the simulated peak from 
$\sim$10 days to $\sim$50 days, for instance, an increase of a factor of $\sim 100$ in the ratio $(E'/E)/(n'/n)$ is needed.
A similar, qualitative analysis was made by \citet{granot03}, who found a similar possible range of values for $B_\parallel/B_\perp$. \citet{stringer20}, on the other hand, presented a statistical study of PD in GRB afterglows. They inferred values of $B_\parallel/B_\perp \approx 0.7-1.3$, consistent with our findings.
Detailed modelling of individual GRBs, or of the entire population of GRBs with observed PD, can potentially put more restrictions on the magnetic field anisotropy \citep[see, e.g.,][]{stringer20, caligula2023}, and is left for future work. 

On August 17, 2017, the Advanced Laser Interferometer Gravitational-wave (GW) Observatory (LIGO) and the Virgo observatory detected the first GW signal from a binary neutron star merger \citep{abbott17}. Approximately $\sim 1.6$~s later, the LIGO signal was followed by a $\gamma$-ray burst (GRB) observed by Fermi and Integral (e.g., \citealt{goldstein17}). The event was observed by numerous telescopes on Earth, covering a wide range of electromagnetic frequencies from radio to X-ray.

\begin{figure}
\includegraphics[width=0.9\columnwidth]{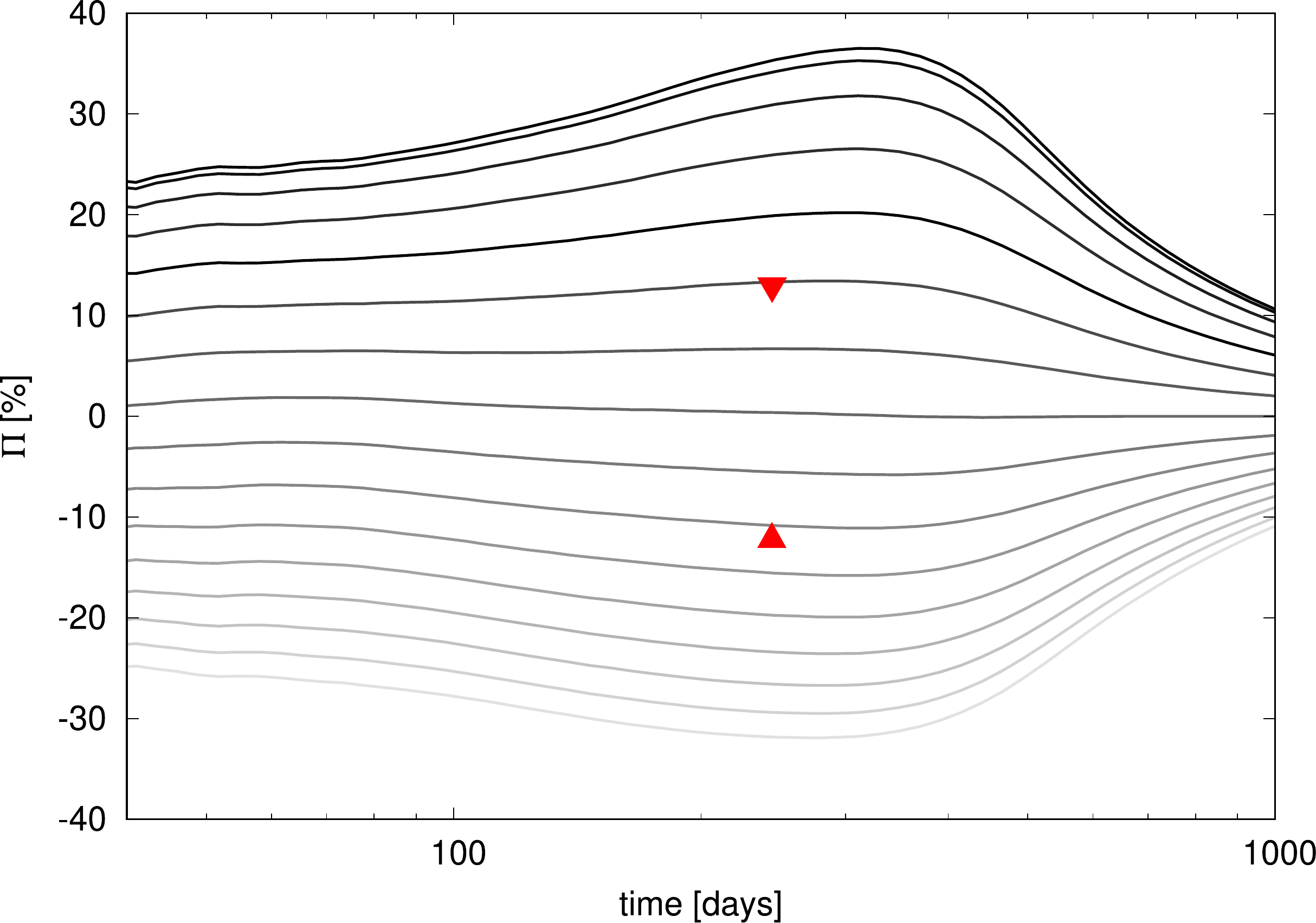}
\caption{PD corresponding to $B_\parallel/B_\perp=0, 0.1, 0.2, 0.3, \dots, 1.5$ (top to bottom curves), and upper limit on the observation of polarisation in GRB 170817a. The following parameters have been used in the numerical calculations: $p = 2.16$, $\nu = 2.8$~GHz, $\epsilon_e=10^{-1.8}$, $\epsilon_B=10^{-3.12}$, $E_{\rm jet}=  10^{50.32}$~erg, $\rho_{\rm amb}=10^{-2}$~cm$^{-3}$, $\theta_{\rm obs}/\theta_{\rm jet} = 3.1$.
}
    \label{fig9}.
\end{figure}

As part of the extensive multi-wavelength follow-up of GRB 170817A, observations by \citet{Corsi18} carried out with the Karl G. Jansky Very Large Array (VLA) obtained an upper limit of 12\% (99\% confidence) for the PD at 244 days in radio frequencies (2.8 GHz). This linear polarisation upper limit has been used by \citet{Corsi18, Gill&Granot2020} to constrain the magnetic field anisotropy in the emitting region. \citet{Gill&Granot2020}, in particular, constrained the magnetic field anisotropy factor to $\eta_B= B_\parallel/B_\perp =0.57-0.89$. Figure \ref{fig9} presents the PD for different values of $\eta_B$, between 0 (upper line) to 1.5 (bottom line), together with the observational upper limit. The value of $\eta$ determined by numerical simulations is $0.5-0.9$, consistent with the value determined above by analysing on-axis GRB polarisation, and the value determined by \citet{Gill&Granot2020}. 

As discussed by \citet{Gill&Granot2020}, this result is not consistent with the magnetic field being only generated by the Weibel two-stream instability, as it would produce mainly a  magnetic field tangled in the shock plane, and it is consistent with a magnetic field  stretched in the post-shock region along the radial direction, which amplify the component of the magnetic field parallel to the velocity. 

Although our simulations considers an initial  top-hat jet (vs. an analytical,  structured jet considered by \citealt{Gill&Granot2020}), we notice that at the time of the polarisation measurement, a top-hat and a Gaussian (structured) jets have a similar structure, which imply that our simulations can be used to constrain the magnetic field anisotropy. On the other hand, the jets at these times are far from being quasi-spherical. This can be seen, e.g., by comparing the PD obtained in our simulations with the one shown by \citet[][figure 2]{Corsi18} for a quasi-spherical ejecta.

\subsection{Limitations and caveats}

The main limitation of this work is that we employed a top-hat structure for the jet. 
Top-hat jets viewed off-axis fail to reproduce the larger X-ray and radio luminosity of GRB 170817A at early times $t<25$ days, and do not naturally account for the observed rise of the non-thermal emission $F_\nu \propto t^{0.8}$ \citep[e.g.,][]{margutti18}, while a steeper dependence, $F_\nu \propto t^{3}$, is expected in top-hat jets.

Nevertheless, we notice that the top-hat phase lasts for a short time. As the jet expands, it acquires a lateral structure \citep{gill19}. This is confirmed by the fact that the PD computed in the previous section shares some similarities with structured jets, differing from top-hat jets computed analytically.   
Also, we notice that we have not considered self-absorption, which can (at early times) be important in radio bands.

The second limitation, shared with previous calculations of light curves and polarisation, is the uncertainty on the post-shock structure of the magnetic field. 
The set of equations integrated into our simulations does not include the evolution of the magnetic field. The magnetic field energy density, indeed, is set by the parameter $\epsilon_B$ as a fraction of the thermal energy, while the geometry of the magnetic field is defined ``by hand''. Actually, the magnetic field intensity can decay in the post-shock region at a different rate with respect to the thermal energy density, i.e. $\epsilon_B$ can change with the distance from the shock. 

The direction of the magnetic field can also change, as noted by \citet{granot03}. As the plasma expands in the post-shock region, the size of each fluid element scales with the self-similarity variable, resulting in a larger stretch in the radial direction than in the tangential direction. This implies that the ratio of the parallel to perpendicular components of the magnetic field can change in the post-shock region, with the component parallel to the velocity being amplified as the fluid element moves away from the shock front. Our simulation also led to an amplification of the magnetic field component parallel to the velocity. To see this, imagine starting with a magnetic field parallel to the shock front in the post-shock region, i.e., perpendicular to the velocity. As the fluid element moves away from the shock front, lateral expansion, instabilities, and turbulence tangle the velocity direction (see figure \ref{fig3}), amplifying the component of the magnetic field parallel to the velocity. We also note that the time- and space-changing magnetic field component perpendicular to the shock front considered in this work is not strictly consistent with the $\nabla \cdot \vec{B}=0$ condition. In fact, the absence of magnetic monopoles $\nabla \cdot \vec{B}=1/r^2 \partial (r^2 B_r)/\partial r = 0$ allows only the (constant in time) radial magnetic field $B_r = A/r^2$ as a solution, which implies that a full multi-dimensional treatment is necessary to establish a physically consistent magnetic field.

\section{Conclusions}

In this paper, we considered a top-hat jet decelerating through a uniform medium. 
We presented the first numerical simulations of linear polarisation during the afterglow phase of a gamma-ray burst. 
We compute the polarisation degree for different magnetic field configurations (parallel and perpendicular to the shock front), for different frequencies, and for observers located on-axis and off-axis with respect to the jet axis.

The behaviour of the on-axis polarisation degree is similar to what was obtained in previous analytical works, while the off-axis PD computed by the numerical simulation differs strongly from the analytical calculation. Instead of two peak observed at intermediate angles ($\theta_{\rm obs}=0.4$ rad), only a single peak, preceded by a shallow increase, is obtained. Furthermore, the peak in the PD drops at larger angles (instead of increasing as observed in analytical calculations). This late time, off-axis behaviour is more consistent with the one typically observed in analytical models of structured jets.

We also computed the magnetic field anisotropy (i.e. the ratio between magnetic field perpendicular and parallel to the shock front) by comparing the numerical model with a sample of on-axis PD observations, and with an upper-limit inferred for the off-axis GRB 170817A. The anisotropy is $\sim 0.5-0.9$, consistent with previous estimate. Our findings emphasise the importance of  capturing accurately the dynamics of the decelerating shock front, in order to properly model future observations of polarisation in off-axis GRB afterglows. This can aid in understanding the structure of the magnetic field in the post-shock region of GRB jets and shed light on its origin.

\section*{Acknowledgements}

We acknowledge the computing time granted by DGTIC UNAM on the supercomputer Miztli (project LANCAD-UNAM-DGTIC-281). We acknowledge the use of chatGPT for English editing of the manuscript. GU acknowledges support from grant 2019/35/B/ST9/04000 from Polish National Science Center.

\section*{Data Availability}

The data underlying this article will be shared on reasonable request to the corresponding author.



\bsp	
\label{lastpage}

\begin{thebibliography}{99}

\bibitem[Abbott et al.(2017)]{abbott17} Abbott, B.~P., Abbott, R., Abbott, T.~D., et al.\ 2017, \apjl, 848, L13 
\bibitem[Aksulu et al.(2020)]{aksulu20} Aksulu, M.~D., Wijers, R.~A.~M.~J., van Eerten, H.~J., et al.\ 2020, \mnras, 497, 4672
\bibitem[Aksulu et al.(2022)]{aksulu22} Aksulu, M.~D., Wijers, R.~A.~M.~J., van Eerten, H.~J., et al.\ 2022, \mnras, 511, 2848
\bibitem[Birenbaum \& Bromberg(2021)]{Birenbaum&Bromberg2021} Birenbaum, G. \& Bromberg, O.\ 2021, \mnras, 506, 4275
\bibitem[Blandford \& McKee(1976)]{blandford76} Blandford, R.~D. \& McKee, C.~F.\ 1976, Physics of Fluids, 19, 1130
\bibitem[Buckley et al.(2021)]{Buckley21} Buckley, D.~A.~H., Bagnulo, S., Britto, R.~J., et al.\ 2021, \mnras, 506, 4621.
\bibitem[Caligula do E.~S. Pedreira et al.(2023)]{caligula2023} Caligula do E.~S. Pedreira, A.~C., Fraija, N., Galvan-Gamez, A., et al.\ 2023, \apj, 942, 81
\bibitem[Cheng et al.(2020)]{cheng20} Cheng, K.~F., Zhao, X.~H., \& Bai, J.~M.\ 2020, \mnras, 498, 3492
\bibitem[Corsi et al.(2018)]{Corsi18} Corsi, A., Hallinan, G.~W., Lazzati, D., et al.\ 2018, \apjl, 861, L10
\bibitem[Covino \& Gotz(2016)]{Covino&Gotz16} Covino, S. \& Gotz, D.\ 2016, Astronomical and Astrophysical Transactions, 29, 205
\bibitem[De Colle et al.(2012)]{decolle12a} De Colle, F., Granot, J., L{\'o}pez-C{\'a}mara, D., et al.\ 2012, \apj, 746, 122
\bibitem[Ghisellini \& Lazzati(1999)]{ghisellini99} Ghisellini, G., \& Lazzati, D.\ 1999, \mnras, 309, L7 
\bibitem[Gill \& Granot(2018)]{gill18} Gill, R., \& Granot, J.\ 2018, \mnras, 478, 4128 
\bibitem[Gill et al.(2019)]{gill19} Gill, R., Granot, J., De Colle, F., et al.\ 2019, \apj, 883, 15
\bibitem[Gill \& Granot(2020)]{Gill&Granot2020} Gill, R. \& Granot, J.\ 2020, \mnras, 491, 5815
\bibitem[Goldstein et al.(2017)]{goldstein17} Goldstein, A., Veres, P., Burns, E., et al.\ 2017, \apjl, 848, L14 
\bibitem[Goodman \& MacFadyen(2008)]{goodman08} Goodman, J. \& MacFadyen, A.\ 2008, Journal of Fluid Mechanics, 604, 325
\bibitem[Granot et al.(2002)]{granot02} Granot, J., Panaitescu, A., Kumar, P., et al.\ 2002, \apjl, 570, L61
\bibitem[Granot \& K{\"o}nigl(2003)]{granot03} Granot, J. \& K{\"o}nigl, A.\ 2003, \apjl, 594, L83
\bibitem[Granot(2003)]{granot03a} Granot, J.\ 2003, \apjl, 596, L17 
\bibitem[Granot(2012)]{granot2012} Granot, J.\ 2012, \mnras, 421, 2610
\bibitem[Gruzinov(1999)]{gruzinov99a} Gruzinov, A.\ 1999, \apjl, 525, L29
\bibitem[Gruzinov \& Waxman(1999)]{gruzinov99b} Gruzinov, A. \& Waxman, E.\ 1999, \apj, 511, 852
\bibitem[Jordana-Mitjans et al.(2020)]{jordana-mitjans20} Jordana-Mitjans, N., Mundell, C.~G., Kobayashi, S., et al.\ 2020, \apj, 892, 97
\bibitem[Klose et al.(2004)]{Klose2004} Klose, S., Palazzi, E., Masetti, N., et al.\ 2004, \aap, 420, 899
\bibitem[Kumar \& Panaitescu(2000)]{kumar00} Kumar, P. \& Panaitescu, A.\ 2000, \apjl, 541, L9
\bibitem[Kumar \& Zhang(2015)]{kumar15} Kumar, P. \& Zhang, B.\ 2015, \physrep, 561, 1
\bibitem[Kuwata et al.(2023)]{kuwata23} Kuwata, A., Toma, K., Kimura, S.~S., et al.\ 2023, \apj, 943, 118
\bibitem[Laing(1980)]{laing80} Laing, R.~A.\ 1980, \mnras, 193, 439 
\bibitem[Lan et al.(2016)]{LanWuDai15} Lan, M.-X., Wu, X.-F., \& Dai, Z.-G.\ 2016, \apj, 826, 128
\bibitem[Lan et al.(2018)]{LanWuDai2018} Lan, M.-X., Wu, X.-F., \& Dai, Z.-G.\ 2018, \apj, 860, 44 
\bibitem[Laskar et al.(2019)]{Laskar19} Laskar, T., Alexander, K.~D., Gill, R., et al.\ 2019, \apjl, 878, L26
\bibitem[Lazzati et al.(2003)]{Lazzati03} Lazzati, D., Covino, S., di Serego Alighieri, S., et al.\ 2003, \aap, 410, 823 
\bibitem[Levan(2018)]{levan18} Levan, A.~J.\ 2018, Gamma-ray bursts, by Levan, Andrew J., 2018. Bristol: IOP Publishing. OCLC: 1082881978. ISBN: 978-0-7503-1500-5; eISBN: 978-0-7503-1502-9
\bibitem[Lyutikov et al.(2003)]{lyutikov03} Lyutikov, M., Pariev, V.~I., \& Blandford, R.~D.\ 2003, \apj, 597, 998
\bibitem[Margutti et al.(2018)]{margutti18} Margutti, R., Alexander, K.~D., Xie, X., et al.\ 2018, \apjl, 856, L18
\bibitem[Medvedev \& Loeb(1999)]{medvedev99} Medvedev, M.~V. \& Loeb, A.\ 1999, \apj, 526, 697
\bibitem[M{\'e}sz{\'a}ros \& Rees(1997)]{meszaros97} M{\'e}sz{\'a}ros, P. \& Rees, M.~J.\ 1997, \apj, 476, 232
\bibitem[Milosavljevi{\'c} \& Nakar(2006)]{milosavljevic06} Milosavljevi{\'c}, M. \& Nakar, E.\ 2006, \apj, 641, 978
\bibitem[Mundell et al.(2013)]{Mundell13} Mundell, C.~G., Kopa{\v{c}}, D., Arnold, D.~M., et al.\ 2013, \nat, 504, 119
\bibitem[Nakar et al.(2003)]{nakar03} Nakar, E., Piran, T., \& Waxman, E.\ 2003, \jcap, 2003, 005
\bibitem[Nava et al.(2016)]{Nava2016} Nava, L., Nakar, E., \& Piran, T.\ 2016, \mnras, 455, 1594
\bibitem[Paczynski \& Rhoads(1993)]{paczynski93} Paczynski, B. \& Rhoads, J.~E.\ 1993, \apjl, 418, L5
\bibitem[Rees \& Meszaros(1992)]{rees92} Rees, M.~J. \& Meszaros, P.\ 1992, \mnras, 258, 41
\bibitem[Rhoads(1997)]{rhoads97} Rhoads, J.~E.\ 1997, \apjl, 487, L1
\bibitem[Rybicki \& Lightman(1979)]{Rybicki&Lightman1979} Rybicki, G.~B. \& Lightman, A.~P.\ 1979, A Wiley-Interscience Publication, New York: Wiley, 1979
\bibitem[Rossi et al.(2004)]{rossi04} Rossi, E.~M., Lazzati, D., Salmonson, J.~D., et al.\ 2004, \mnras, 354, 86
\bibitem[Sari et al.(1998)]{sari98} Sari, R., Piran, T., \& Narayan, R.\ 1998, \apjl, 497, L17
\bibitem[Sari(1999)]{sari99a} Sari, R.\ 1999, \apjl, 524, L43. 
\bibitem[Sari et al.(1999)]{sari99} Sari, R., Piran, T., \& Halpern, J.~P.\ 1999, \apjl, 519, L17
\bibitem[Shimoda \& Toma(2021)]{Shimoda&Toma2021} Shimoda, J. \& Toma, K.\ 2021, \apj, 913, 58
\bibitem[Sironi \& Goodman(2007)]{sironi07} Sironi, L. \& Goodman, J.\ 2007, \apj, 671, 1858
\bibitem[Stringer \& Lazzati(2020)]{stringer20} Stringer, E. \& Lazzati, D.\ 2020, \apj, 892, 131. doi:10.3847/1538-4357/ab76d2
\bibitem[Teboul \& Shaviv(2021)]{Teboul&Shaviv2021} Teboul, O. \& Shaviv, N.~J.\ 2021, \mnras, 507, 5340
\bibitem[Toma et al.(2008)]{Toma2008} Toma, K., Ioka, K., \& Nakamura, T.\ 2008, \apjl, 673, L123
\bibitem[Urata et al.(2019)]{Urata19} Urata, Y., Toma, K., Huang, K., et al.\ 2019, \apjl, 884, L58
\bibitem[van Eerten \& MacFadyen(2012)]{vaneerten&macfadyen2012} van Eerten, H.~J. \& MacFadyen, A.~I.\ 2012, \apjl, 747, L30
\bibitem[Wu et al.(2005)]{wu05} Wu, X.~F., Dai, Z.~G., Huang, Y.~F., et al.\ 2005, \mnras, 357, 1197. doi:10.1111/j.1365-2966.2005.08685.x






\end{thebibliography}
\end{document}